\documentclass[usegraphicx]{mn2e}
\usepackage{subfig}
\usepackage{epsfig}
\usepackage{amsmath}
\usepackage{amssymb}
\usepackage{graphicx}
\usepackage{times,longtable,color}

\usepackage{wrapfig}

\newcommand {\apgt} {\ {\raise-.5ex\hbox{$\buildrel>\over\sim$}}\ }
\newcommand {\aplt} {\ {\raise-.5ex\hbox{$\buildrel<\over\sim$}}\ } 
\newcommand {\degree}{$^{\circ}$}

\title[LT precession as a means to constrain the EoS]
{Lense-Thirring precession in ULXs as a possible means to constrain the neutron star equation-of-state}

\author[M. Middleton et al.]
{M. J. Middleton$^{1}$, P. C. Fragile$^{2}$, M. Bachetti$^{3}$, M. Brightman$^{4}$, Y-F. Jiang$^{5}$, \newauthor W. C. G. Ho$^{1,6}$,  T. P. Roberts$^{7}$, A. R. Ingram$^{8}$, T. Dauser$^{9}$, C. Pinto$^{10}$, D. J. Walton$^{10}$,  \newauthor F. Fuerst$^{11}$, A. C. Fabian$^{10}$ \& N. Gehrels$^{12}$\\
\\
1. Department of Physics and Astronomy, University of Southampton, Highfield, Southampton SO17 1BJ, UK\\
2. Department of Physics and Astronomy, College of Charleston, Charleston, SC 29424, USA\\
3. INAF/Osservatorio Astronomico di Cagliari, via della Scienza 5, I-09047 Selargius (CA), Italy\\
4. Cahill Center for Astrophysics, California Institute of Technology, 1216 East California Boulevard, Pasadena, CA 91125, USA\\
5. Kavli Institute of Theoretical Physics, University of California, Santa Barbara, CA 93110, USA\\
6. Mathematical Sciences and STAG Research Centre, University of Southampton, Southampton SO17 1BJ, UK\\
7. Centre for Extragalactic Astronomy, Durham University, Dept of Physics, South Road, Durham DH1 3LE, UK\\
8. Anton Pannekoek Institute for Astronomy, University of Amsterdam, Science Park 904, NL-1098 XH Amsterdam, the Netherlands\\
9. Remeis Observatory \& ECAP, Universit{\"a}t Erlangen-N{\"u}rnberg, Sternwartstr. 7, 96049, Bamberg, Germany\\
10. Institute for Astronomy, University of Cambridge, Cambridge, UK\\
11. European Space Astronomy Centre (ESA/ESAC), Science Operations Department, Villanueva de la Canada (Madrid), Spain\\
12. NASA Goddard Space Flight Center, Mail Code 661, Greenbelt, MD 20771, USA\\
}

\pagerange{\pageref{firstpage}--\pageref{lastpage}} \pubyear{2017}
\long\def\symbolfootnote[#1]#2{\begingroup\def\thefootnote{\fnsymbol{footnote}}\footnote[#1]{#2}\endgroup} 
\def\ga{\mathrel{\hbox{\rlap{\hbox{\lower4pt\hbox{$\sim$}}}{\raise2pt\hbox{$>$}}
}}}

\begin{document}

\topmargin = -0.5cm

\maketitle

\label{firstpage}

\begin{abstract}

The presence of neutron stars in at least three ultraluminous X-ray sources is now firmly established and offers an unambiguous view of super-critical accretion. All three systems show long-timescale periods (60-80 days) in the X-rays and/or optical, two of which are known to be super-orbital in nature. Should the flow be classically super critical, i.e. the Eddington limit is reached locally in the disc (implying surface dipole fields that are sub-magnetar in strength), then the large scale-height flow can precess through the Lense-Thirring effect which could provide an explanation for the observed super-orbital periods. By connecting the details of the Lense-Thirring effect with the observed pulsar spin period, we are able to infer the moment-of-inertia and therefore equation-of-state of the neutron star without relying on the inclination of, or distance to the system. We apply our technique to the case of NGC 7793 P13 and demonstrate that stronger magnetic fields imply stiffer equations of state. We discuss the caveats and uncertainties, many of which can be addressed through forthcoming radiative magnetohydrodynamic (RMHD) simulations and their connection to observation. 

\end{abstract}


\section{introduction}

The majority of ultraluminous X-ray sources (ULXs - see Kaaret, Feng \& Roberts 2017) are now thought to provide a window into the nature of super-critical accretion onto stellar remnants (Shakura \& Sunyaev 1973). Direct insights can be obtained via the shape of the X-ray spectra (Stobbart et al. 2006; Gladstone et al. 2009; Poutanen et a. 2007), coupled  spectral-timing evolution (Sutton et al. 2013; Middleton et al. 2015a), and mass and energy loss in the equatorial winds (Middleton et al. 2014; 2015b; Pinto, Middleton \& Fabian 2016; 2017; Walton et al. 2016a) and via jets (Cseh et al. 2014, 2015; Miller-Jones et al. in prep). This information can then be directly compared to the outputs of 3D RMHD simulations (Ohsuga et al. 2009; Jiang, Stone \& Davis 2014; S{\c a}dowski et al. 2014; 2016). As theory and observation begin to converge, there is hope that we will obtain a better understanding of the growth of the most massive quasars at very high redshift (Fan et al. 2003; Wu et al. 2015) which are required to have grown at super-critical rates (e.g. Volonteri et al. 2005), and of tidal disruption events (e.g. Begelman \& Volonteri 2016). With the recent discovery that a number of ULXs host neutron stars (Bachetti et al. 2014; Fuerst et a. 2016; Israel et al. 2017a; 2017b), new possibilities have arisen, including the possibility of constraining the equation of state (EoS) in these ultraluminous pulsars (ULPs).

Determining the neutron star EoS is a major goal of observational astronomers and theoretical nuclear physicists alike (see the review of Lattimer \& Prakash 2016) as exemplified by NASA's new NICER mission (see Gendreau et al. 2016). From an observational standpoint, studying the photospheric radius expansion (PRE) subset of type I X-ray bursts can allow for simultaneous mass and radius measurements (Steiner et al. 2010; Nattila et al. 2016) although this typically requires that the photospheric radius be the same as that of the stellar radius which may not be correct. Arguably, the most promising means of obtaining constraints on the EoS relies on determining the radius of the neutron star via the cool surface emission in quiescent LMXBs (Heinke et al. 2003, 2006; Webb \& Barret 2007) and obtaining corresponding mass estimates via independent means. Whilst the largest uncertainty in this technique is potentially the distance estimate (whilst source variability, absorption by interstellar hydrogen and the effect of the neutron star atmosphere are secondary effects), this can be significantly improved upon by studying sources in globular clusters where the distance is known in some cases to 3\% accuracy (Heinke et al. 2014). Here we explore the implications for obtaining estimates of the EoS via a new approach: Lense-Thirring precession of the wind in ULPs which, whilst dependent on a number of observational quantities, does not depend on distance nor inclination (which can often be hard to measure accurately).

\section{Lense-Thirring precession as a means to constrain the equation-of-state}

Lense-Thirring precession (see Bardeen \& Petterson 1975) is a general relativistic effect which occurs when an orbiting particle (or mass of particles in a fluid) is displaced vertically from a rotating body's equatorial axis such that frame-dragging then induces oscillations about the ecliptic and periapsis. This has recently been shown to occur in the weak-field limit around the Earth by the Gravity Probe B experiment (Everitt et al. 2015) and in the strong-field limit in black hole binaries (BHBs) at accretion rates below the Eddington limit, where the relatively large scale-height corona of hot thermal electrons in the inner regions (Ichimaru 1977; Esin, McClintock \& Narayan 1997; Poutanen, Krolik \& Ryde 1997) precesses as a solid body (Fragile et al. 2007). In so-doing the Lense-Thirring  precession of the corona self-consistently explains the ubiquitous low frequency quasi-periodic oscillations (QPOs; see Stella \& Vietri 1998; Ingram et al. 2009, 2012, 2016, 2017), where the characteristic QPO timescale depends on the outer radius of the corona (i.e. the timescale of precession increases with increasing truncation radius, see Done et al. 2007). This mechanism fundamentally requires the rotation axis of the compact object to be misaligned with that of the binary orbit, although this is expected to be a common outcome of a supernova (Fragos et al. 2010). As the Lense-Thirring torque in this regime (where the viscosity parameter, $\alpha$, is less than the scale-height of the flow) is communicated via bending waves propagating outwards at the gas sound speed, a further key requirement is that the sound crossing time in the flow is shorter than the precession timescale (Fragile et al. 2007). Should the various requirements be met at accretion rates exceeding the Eddington limit, there is no theoretical reason why Lense-Thirring  precession should not also occur in this regime.

When the accretion rate exceeds the Eddington limit (we say it is super-critical: Shakura \& Sunyaev 1973) the structure of the accretion flow changes at the spherisation radius ($r_{sph}$) due to the intense radiation pressure. Within this radius, the scale-height ($H/R$ - where $H$ is the height of the disc from the mid-plane at a radius $R$) of the flow tends to unity, which it retains down to the innermost edge of the accretion disc (Poutanen et al. 2007). In the classical picture, the flow cools via advection (Abramowicz et al.1988) and the launching of winds from the surface of the disc (Poutanen et al. 2007). Whilst the latest 3D RMHD simulations (Jiang et al. 2014; S{\c a}dowski et al. 2014) have confirmed much of this classical picture, they also reveal that vertical advection of flux due to magnetic buoyancy acts to increase the radiative efficiency (thought to be very low in super-critical flows) even in the presence of radial advection. 
 
In super-critical flows, $H/R > \alpha$ out to $r_{sph}$, and this entire region can therefore precess as a solid body. As the characteristic radius is large (and sound crossing time is long), the precession timescale is correspondingly long. As presented in Poutanen et al. (2007), the position of $r_{sph}$ is given by:

\begin{equation}
\frac{r_{sph}}{r_{isco}\dot{m}} \approx 1.34-0.4\epsilon_{wind} + 0.1\epsilon_{wind}^{2} - \left(1.1-0.7\epsilon_{wind}\right)\dot{m}^{-2/3}
\end{equation}

\noindent where all radii are in units of the gravitational radius ($R_{g} = GM/c^{2}$ where $M$ is the mass of the compact object) and $r_{isco}$ is assumed to be the radius of the innermost stable circular orbit (ISCO, and therefore depends on the compact object's angular momentum). In the above formula, $\dot{m}$ is the mass accretion rate through the outer disc in units of the Eddington mass accretion rate, i.e. $\dot{m} = \dot{M}/\dot{M}_{Edd}$ where $\dot{M}_{Edd} = L_{Edd}/\eta c^2$, $L_{Edd} \approx 1.26\times10^{38}M/M_{\odot}$ erg/s and $\eta$ is the radiative efficiency (which in the case of super-critical flows is thought to be low and we assume $\eta$ = 1\% throughout unless stated otherwise), such that in super-critical accretion, $\dot{m} >$ 1. Finally, $\epsilon_{wind}$ is the fraction of radiative energy spent in launching the wind relative to that observed and can be determined observationally from $\epsilon_{wind} = (1+L_{rad}/L_{wind})^{-1}$ where $L_{rad}$ is the observed radiative luminosity and $L_{wind}$ is the kinetic luminosity of the wind. In the super-critical model of Poutanen et al. (2007),  $\dot{m}$ is related to the observed (colour) temperature of the quasi-blackbody emission at $r_{sph}$:
 
\begin{equation}
T_{sph} \approx 1.5 f_{col}\dot{m}^{-1/2}m^{-1/4} [keV]
\end{equation}

In the above formula, $m = M/M_{\odot}$ and $f_{col}$ is the colour temperature correction factor which accounts for the scattering and absorption of photons escaping from the inflow and which modifies the intrinsic temperature ($T_{int}$) to what we instead observe ($T_{sph}$). For electron scattering opacity alone, $f_{col}$ is often taken to be $\sim$1.7 (Shimura \& Takahara 1995), however, an accurate determination of this value must account for {\it both} absorption and scattering opacities and theoretically saturates at $\sim (72/T_{int})^{1/9}$ (Davis et al. 2006). In the spectrum of ULXs, the emission at soft X-ray energies is thought to be associated with the spherisation radius (Poutanen et al. 2007; Middleton et al. 2015a) allowing an estimate of the combination of mass and accretion rate to be obtained with relative ease.  

The precession timescale of the disc at $r_{sph}$ depends on the surface density profile of the inflow which is commonly assumed to go as $\Sigma \propto r^{-\xi}$ - in the case of super-critical discs, $\xi$ can be obtained analytically from standard super-critical disc theory incorporating mass loss (Shakura \& Sunyaev 1973; Poutanen et al. 2007) which indicates that $\dot{m} \propto r$. The surface density is related to $\dot{m}$ by $\dot{m} = \Sigma 2\pi rv_{r}$ where $v_r$ is the radial infall velocity; assuming viscous infall, $v_{r} = r/t_{visc}$ where $t_{visc}$ is the viscous timescale and is proportional to the dynamical timescale ($t_{dyn}$). As $t_{dyn} \propto r^{3/2}$ we then find that $\Sigma \propto r^{1/2}$, i.e. $\xi$ = -0.5. From the formula of Fragile et al. (2007) we then obtain the formula for the precession period ($P_{prec}$) at $r_{sph}$:

\begin{equation}
P_{prec} = \frac{GM\pi}{3c^{3}a_{*}} r_{sph}^{3} \left[ \frac{1-\left(\frac{r_{in}}{r_{sph}}\right)^{3}}{ln\left(\frac{r_{sph}}{r_{in}}\right)}\right] [s]
\end{equation}

\noindent where $a_{*}$ is the dimensionless spin value (= $Jc/GM^{2}$ where $J$ is the angular momentum of the compact object) and $r_{in}$ is the inner edge of the disc. Whilst this is the precession period associated with the {\it inflow}, radiatively driven outflows launched from the disc should conserve angular momentum and remain optically thick (and therefore present an effective obscuring envelope) out to a radius at which they become optically thin. Poutanen et al. (2007) determine this `photospheric' radius to be:

\begin{equation}
r_{out} \approx \frac{3\epsilon_{wind}}{\psi\phi}\dot{m}^{3/2}r_{isco}
\end{equation}

\noindent where $\psi$ is the the ratio of asymptotic wind velocity ($v_{wind}$) to the Keplerian velocity at $r_{sph}$ (and so is a function of both $\dot{m}$ and $a_{*}$) and $\phi$ is the cotangent of the opening angle of the wind cone. Accounting for conservation of angular momentum then allows us to obtain the precession period of the optically thick envelope:

\begin{equation}
P_{prec} = \frac{GM\pi}{3c^{3}a_{*}} r_{sph}^{3} \left[ \frac{1-\left(\frac{r_{in}}{r_{sph}}\right)^{3}}{ln\left(\frac{r_{sph}}{r_{in}}\right)}\right] \times \left(\frac{r_{out}}{r_{sph}}\right)^{2} [s]
\end{equation}

By inspection of the above, it is clear that Lense-Thirring precession, if occurring in ULXs could be a powerful tool to determine bounds on the mass and spin parameter space which is especially relevant when the primary is a black hole (due to the lack of complicating surface effects). When the compact object is a neutron star, we may also have a spin period ($P_{spin}$) revealed via pulsations which is connected to both $a_*$ and - crucially - the neutron star moment-of-inertia ($I$) by:

\begin{equation}
a_{*} = \frac{2\pi Ic}{P_{spin}GM^{2}}
\end{equation}

\noindent We therefore have a series of steps to estimate the moment-of-inertia (which is connected to the EoS) for a ULX hosting a neutron star: 

\begin{enumerate}
\item{estimate $\dot{m}$ from the X-ray spectrum (equation 2) for a given compact object mass}
\item{from $\dot{m}$ and equations 1 \& 4, estimate $r_{sph}$ and $r_{out}$ for a given set of physical wind values (i.e. $\epsilon_{wind}$, $v_{wind}$ and $\phi$) which we discuss in \S3.2 for one ULP and should be considered on a case-by-case basis.}
\item{for an observed $P_{prec}$, obtain a range in $a_{*}$ using equation 5 and the mass in (i)}
\item{for each mass we then have a corresponding value of $a_{*}$ which allows us to estimate $I$ from equation 6 for an observed $P_{spin}$}
\end{enumerate}

Although there are a number of sources of uncertainty in this approach (as we discuss in \S4), it is notable that neither distance nor inclination are present in the above formulae. Finally, we note that the method is applicable only to black holes and neutron stars with low-to-moderate surface dipole field strengths which we discuss further in  \S3.

\section{Application to the ultraluminous pulsar NGC 7793 P13}

To-date there are three reported ULPs: M82 X-2 (P$_{spin}$ = 1.37~s,  Bachetti et al. 2014), NGC 5907 ULX1 (P$_{spin}$ = 1.13~s, Israel et al. 2017a) and NGC 7793 P13 (P$_{spin}$ = 0.42~s, Fuerst et al. 2016; Israel et al. 2017b). Both NGC 5907 ULX1 and NGC 7793 P13 (referred to as P13 hereafter) have a well constrained period in the X-rays and/or optical bands lasting 60-80 days (Motch et al. 2014; Walton et al. 2016b; Hu et al. 2017) whilst M82 X-2 has a reported period of $\sim$62 days (Kaaret et al. 2006; Pasham \& Strohmayer 2013; Kong et al. 2016; Brightman et al. submitted). These periods have been interpreted as orbital or super-orbital in nature, however, a shorter (day timescale) period in M82 X-2 has been identified with the binary orbit (Bachetti et al. 2014), consistent with the presence of a high mass companion star. In the case of P13, it is argued that the He II line - which is claimed to follow the optical periodicity of 63.52 days (Motch et al. 2014) - cannot be from the secondary star (Fabrika et al. 2015) and must instead be associated with a wind from a super-critical accretion disc and therefore the periodicity is not {\it necessarily} associated with the orbit of the companion star. Of the three ULPs, P13 is by far the easiest to study given its high X-ray flux and ease with which it can be spatially resolved by X-ray telescopes (i.e. it is not confused with nearby sources), we therefore apply our new technique to this source to explore possible constraints on the EoS assuming Lense-Thirring precession drives the 63.52 day period.

\subsection{X-ray spectroscopy}

X-ray observations of NGC 7793 P13 were obtained in November 2013 by ESA's {\it XMM-Newton}. We extract the spectral products for all three EPIC cameras (PN, MOS1 and MOS2) using SAS v 15.0 and 35 arcsec extraction regions following standard procedures as outlined in the user's manual\footnotemark\footnotetext{https://heasarc.gsfc.nasa.gov/docs/xmm/sas/USG/} after subtracting periods of soft proton flares in the high energy ($>$10 keV) background. Subsequent spectral fitting was performed on re-binned data (to have 20 counts/spectral bin for chi-squared fitting) using the {\sc xspec} package (Arnaud et al. 1996).

As shown in Middleton et al. (2015a), the X-ray spectra of ULXs can be well described by the combination of quasi-thermal disc emission (parameterised using the model {\sc diskbb}: Mitsuda et al. 1984) and thermal Compton up-scattering (parameterised using the model {\sc nthcomp}: Zdziarski et al. 1996; {\.Z}ycki, Done \& Smith 1999) with the seed photons tied to those of the {\sc diskbb} component (see Middleton et al. 2015a for the physical reasoning behind this). We apply these model components to the spectral data of P13 and account for neutral absorption using {\sc tbabs} with appropriate abundance tables (Wilms et al. 2000) and lower limit set to the Galactic column in the direction of NGC 7793 (1.2$\times$10$^{20}$ cm$^{-2}$: Dickey \& Lockman 1990). Finally, a multiplicative constant is added to account for any differences in the response of the instruments. The statistical quality of the resulting fit, shown in Figure 1 (and re-binned for clarity), is very good: 653/703 degrees of freedom (with the parameters and 1 $\sigma$ errors shown in Table 1). Notably the temperature of the soft component is very well constrained - this is important as $T_{sph}$ enters into our calculation of the spherisation radius (via $\dot{m}$ - equation 2). The cross-normalisation across all three instruments is consistent with unity to within 5\%.

As shown in Figure 1, residuals to the best-fitting spectral model of P13 indicate the same {\it overall} pattern as seen in other ULXs at soft energies (Middleton et al. 2015b) and a fit with a constant (set at unity) to these over the 0.5--2 keV energy range is statistically excluded at $>$ 2 $\sigma$ (null hypothesis probability $<$ 0.05). Such residuals to the best-fitting continuum models of ULXs have been reported for many years and were initially interpreted as emission lines from collisionally excited plasma associated with star formation local to the ULX. We now know from high resolution imaging of an archetypal ULX that it cannot be associated with gas beyond $\sim$25pc of the source (Sutton et al. 2015) and in Middleton et al. (2014) the residuals were re-interpreted as low resolution, blue-shifted atomic features from an outflowing wind for the first time. This was subsequently confirmed in a high energy-resolution spectral analysis using {\it XMM-Newton}'s RGS at soft energies (Pinto et al. 2016, 2017) and a CCD-energy-resolution spectral analysis at higher energies in a combined {\it XMM-Newton/NuSTAR} study (Walton et al. 2016a). The features at low (CCD-quality) energy-resolution appear ubiquitous in the population of bright ULXs and in one well studied case decrease in strength with spectral hardness (Middleton et al. 2015b), thought to indicate an equatorial (i.e. non-polar) wind geometry. Although high energy-resolution data is not yet available to confirm our assertion that the residuals to the best-fitting model for P13 are due to an outflow, given the firm association with atomic features in other ULXs, we take this as tentative evidence that winds are also present in this source, concordant with the He II lines seen in the optical (Motch et al. 2014). 

Associating the spectral residuals in P13 with unresolved atomic lines in emission and absorption resulting from an outflow (Middleton et al. 2014, 2016; Pinto et al. 2016; Walton et al. 2016a) also allows us to explain the periodic brightening in the optical by precession of the wind cone. Although the the companion star is always illuminated by the accretion flow onto the neutron star, we should only see this for orbital phases when the latter is approaching inferior conjunction, with the brightest occurrence when the wind cone is tilted towards the companion star. In this picture, the reprocessed optical emission would peak on the timescale of the precession whilst the orbital period would appear as a modulation on shorter timescales. This has the attraction that it can explain why the He II lines (associated with the wind: Fabrika et al. 2015) are out of phase with the optical brightness (Motch et al. 2014): the observed line velocity is a function of the inclination of the wind to the observer (e.g. maximum blue-shifted when the wind travels directly towards us) and so the chance association of the maximal illumination of the secondary with maximum observed wind velocity is small. Such a mechanism would also predict that any modulation in the X-rays be out of phase with the optical modulation as appears to be the case (Motch et al. 2014). Finally we note that precession of a wind-cone may {\it naturally} lead to increased optical emission on the precession timescale regardless of illuminating the secondary star due to increased low energy emission at more edge-on (to the wind) phases due to scattering through the optically thick wind/inflow - again, this would predict out-of-phase correlations between the optical brightness, optical lines and X-ray emission. In future we will investigate the impact of the system parameters on the correlations between different bands but for now we restrict ourselves to a discussion of the implications of the observed super-orbital periods being driven by Lense-Thirring precession.

\begin{figure}
\begin{center}
\includegraphics[trim = 70 0 0 0, width=9cm]{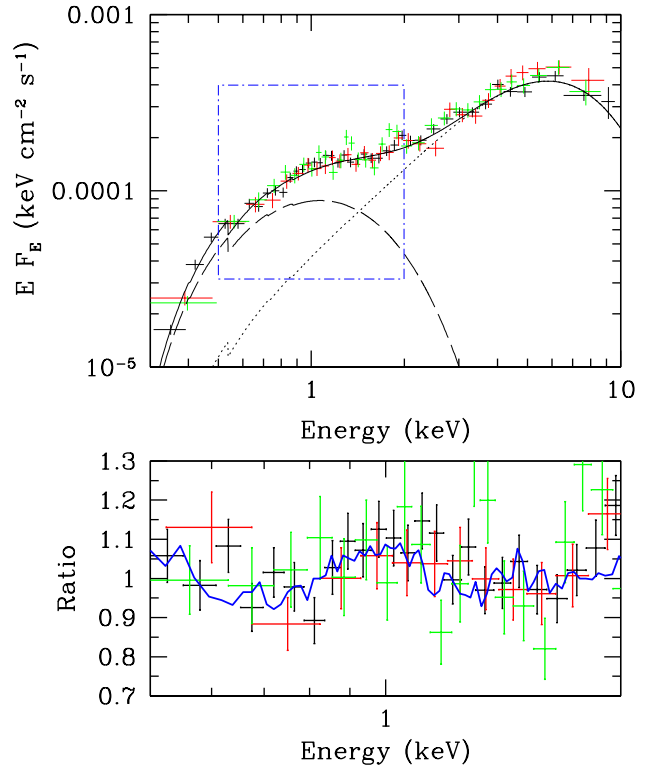}
\end{center}
\vspace{-0.2cm}
\caption{{\it Upper panel}: {\it XMM-Newton} spectral data points (PN: black, MOS1: red and MOS2: green) for P13 (re-binned for clarity). The best-fitting, unfolded model of {\sc tbabs*(diskbb+nthcomp)} is shown as a solid line (see Table 1 for parameters and 1 $\sigma$ errors) with the {\sc diskbb} component shown as a dashed line and {\sc nthcomp} as a dotted line. The blue, dot-dashed bounding box indicates the region of interest for the study of residuals to the model. {\it Lower panel}: 0.5-2 keV residuals to the best-fitting spectral model (same colour scheme as the upper panel). These cannot be adequately described by a constant at unity (null hypothesis probability $<$ 0.05) but resemble the residuals seen in other ULXs (Middleton et al. 2015b). For reference, the blue line indicates the smoothed residuals of the ULX NGC 1313 X-1 (Middleton et al. 2015b) which have since been unambiguously resolved into emission and absorption features associated with a relativistically outflowing wind (Pinto et al. 2016; Walton et al. 2016a).}
\label{fig:l}
\end{figure}

\begin{table}
\begin{center}
\begin{minipage}{60mm}
\bigskip
\caption{Spectral fitting results}
\begin{tabular}{|c|c|c|c|c|c|c|c}
  \hline
 \multicolumn{2}{c}{\sc tbabs*(diskbb+nthcomp)} \\
\hline
 Model parameter & value (1 $\sigma$ error) \\
   \hline
nH ($\times$10$^{20}$cm$^{-2})$ & 8.5 $\pm$ 0.4\\
kT$_{in}$ (keV) & 0.385 $\pm$ 0.003\\ 
norm {\sc diskbb} & 0.69 $\pm$ 0.01\\
$\Gamma$ & 1.215 $\pm$ 0.004\\
kT$_{e}$ (keV) & 1.57 $\pm$ 0.10\\
norm {\sc nthcomp} ($\times$10$^{ -5}$)& 4.88 $\pm$ 0.07\\
$\chi^{2}$/d.o.f   & 653/703\\
\hline
\end{tabular}
Notes: Best-fitting model parameters for the fit to the PN, MOS1 and MOS2 data of NGC 7793 P13 (see Figure 1). Errors are quoted at 1 $\sigma$.

\end{minipage} 

\end{center}
\end{table}

\begin{figure}
\begin{center}
\includegraphics[trim=80 120 20 20, clip, width=8cm]{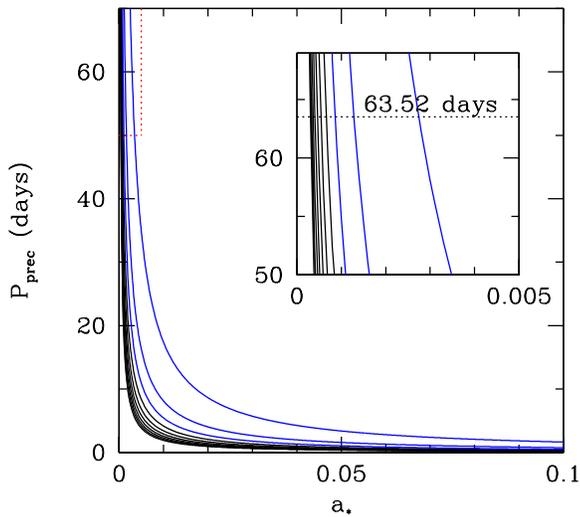}
\end{center}
\vspace{-0.2cm}
\caption{Precession period, $P_{prec}$, versus dimensionless spin parameter $a_{*}$. Using estimated wind parameters and $T_{sph}$ from the X-ray spectrum as input (see Table 1), we determine the predicted precession period for a given compact object mass and spin. Masses shown in black are typical black hole masses (3-10 M$_{\odot}$ in steps of 1 M$_{\odot}$) and in blue are neutron star masses (1-3 M$_{\odot}$ in steps of 1 M$_{\odot}$). The inset shows a zoom-in of the region bound by the red dotted lines which allows the range of mass-spin values for the observed precession period of P13 (Motch et al. 2014, indicated by a horizontal dashed line) to be more clearly discerned.}
\label{fig:l}
\end{figure}

\subsection{Mass-spin constraints}

\begin{table*}
\begin{center}
\begin{minipage}{120mm}
\bigskip
\caption{Example model parameters}
\begin{tabular}{c|ccc|ccc|ccc}
  \hline
 
 Model parameter & \multicolumn{3}{c} {1 M$_{\odot}$} &  \multicolumn{3}{c} {1.5 M$_{\odot}$}  &  \multicolumn{3}{c} {2 M$_{\odot}$}\\
 \hline
 Dipole Field strength (G) & 10$^{10}$ & 10$^{11}$ & 10$^{12}$ & 10$^{10}$ & 10$^{11}$ & 10$^{12}$ & 10$^{10}$ & 10$^{11}$ & 10$^{12}$ \\
 $\dot{m}$ ($\dot{M}_{Edd}$) & \multicolumn{3}{c} {49} & \multicolumn{3}{c} {40} & \multicolumn{3}{c}  {34} \\
 $r_{sph}$ ($R_{g}$)  & \multicolumn{3}{c}  {323} & \multicolumn{3}{c} {262} & \multicolumn{3}{c} {225}\\
 $r_{out}$  ($R_{g}$) &  \multicolumn{3}{c}  {2419} & \multicolumn{3}{c} {1983} & \multicolumn{3}{c} {1722} \\
 $r_{m}$ ($R_{g}$) & 25 & 70 & 195 & 16 & 45 & 124 & 12 & 32 & 90 \\
 $P_{prec}$ (days)  & 44 & 73 & 175 & 33 & 52 & 111 & 27 & 41 & 82 \\
\hline
\end{tabular}
Notes: Example model parameters for obtaining the precession period of the wind for a range of neutron star masses and dipole field strengths assuming $a_{*}$ = 0.001, $f_{col} \approx$ 1.79, $\epsilon_{wind}$ = 0.5, $v_{wind}$ = 0.1c and $\phi$ = 0.7.

\end{minipage} 

\end{center}
\end{table*}

From the X-ray spectral fitting described above, we have a best-fitting temperature for the soft X-ray component of 0.385~keV; from this we can determine $\dot{m}$ for a given mass (equation 2) by assuming the soft component traces the spherisation radius (Poutanen et al. 2007). This requires an estimate for the colour temperature correction factor ($f_{col}$); whilst the actual value must depend on as-yet-unknown details of the accretion flow, we should expect an approximate {\it lower limit} to be given by $(72/T_{sph})^{1/9}$ which is $\approx$ 1.79 and we use this value throughout, investigating the impact of uncertainties in \S 4.3 (noting that improved values for $f_{col}$ may be possible in future).

We can estimate $\dot{m}$ and the positions of $r_{sph}$ and $r_{out}$ (for a given mass) using equations 1, 2 \& 4, and thereby calculate a {\it predicted} precession period for a given compact object mass and $a_*$. This also requires estimates for the physical parameters of the wind, namely $\psi$, $\phi$ and $\epsilon_{wind}$. $\psi$ is determined from the ratio of the wind velocity ($v_{wind}$) to the Keplerian velocity at $r_{sph}$ for each mass and accretion rate; given that we detect low energy-resolution line features (Figure 1), similar in overall shape and energy to those seen in other ULXs (see Middleton et al. 2014; 2015b) we assume $v_{wind} \approx$ 0.1c to be broadly consistent with modelling of the features in these other sources (Middleton et al. 2014; 2015b; Pinto et al. 2016; 2017) and discuss the impact of different wind speeds in \S4.5. We assume the opening angle of the wind cone in P13 to be consistent with predictions from 3D RMHD simulations where the opening angle is $\approx$55\degree and does not change substantially until $\dot{m} \approx$ 200; (which is above the range of mass accretion rates inferred here: $\dot{m} \lesssim$ 50); this gives an estimate for $\phi$ of 0.7 which we discuss further in \S4.7.   

It is possible to estimate the kinetic luminosity of the wind from the absorption lines (Pinto et al. 2016), however, this requires a value for the ionising luminosity which is non-trivial to obtain. Instead we assume that $\epsilon_{wind}$ is limited by the approximate theoretical values found in 3D RMHD simulations of $\approx$ 0.25-0.5 (Jiang et al. 2014; S{\c a}dowski et al. 2016 and depends on mass accretion rate) and assume $\epsilon_{wind}$ = 0.5 in the following analysis, addressing the uncertainty in this value in \S4.6.

Given that we detect pulsations in P13 ($P_{spin}$ = 0.42 s: Fuerst et al. 2016) it is safe to assume that the inner edge of the disc does not reside at the star's surface and instead must lie around the magnetospheric radius ($R_{m}$) which, for a thin disc, depends on the dipole field strength, accretion rate and neutron star mass according to the commonly used formula  (Davidson \& Ostriker 1973):

\begin{figure*}
\begin{center}
\includegraphics[trim=90 90 40 40, clip, width=12cm]{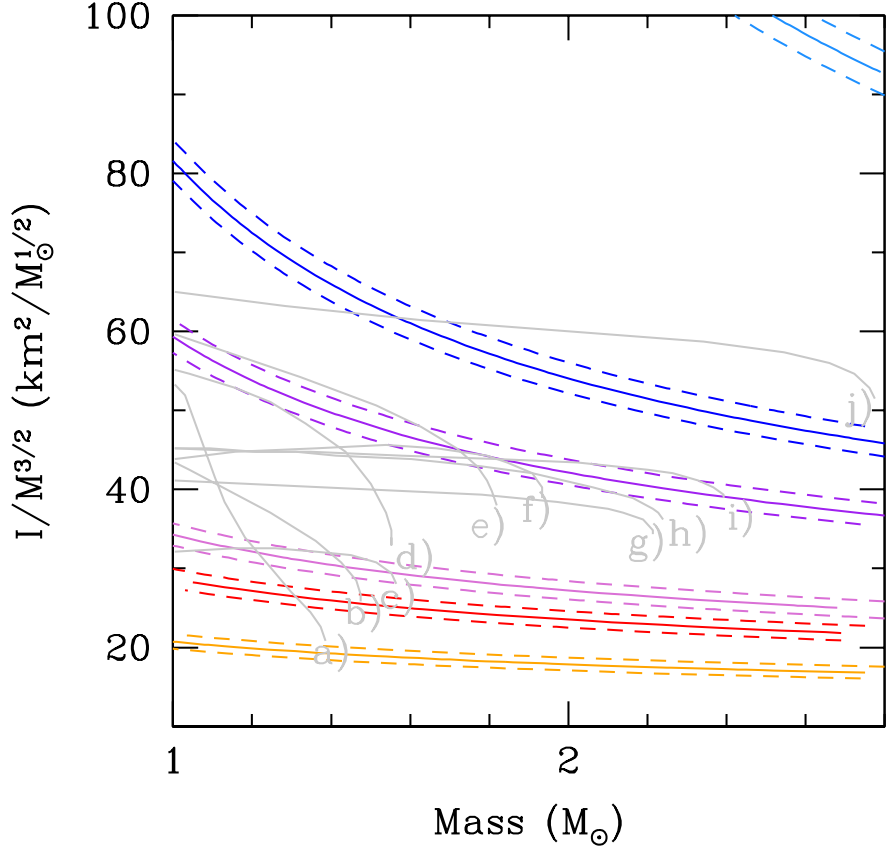}
\end{center}
\vspace{-0.2cm}
\caption{Moment-of-inertia (normalized by $M^{3/2}$) versus mass for NGC 7793 P13 for a range of surface dipole field strengths: B = 1$\times$10$^{10}$ G (orange), 5$\times$10$^{10}$ G (red), 1$\times$10$^{11}$ G (pink), 5$\times$10$^{11}$ G (purple), 1$\times$10$^{12}$ G (blue) and 5$\times$10$^{12}$ G (light blue). In each case, the dashed lines indicate the 1 $\sigma$ error on $T_{sph}$ propagated into $r_{sph}$ and $r_{out}$ (and our limiting value of $f_{col}$). The EoS (a) to j)) are taken from Lattimer \&  Schutz (2005) and correspond to a) GS1, b) PAL6, c) SQM1, d) GM3, e) MS1, f) SQM3, g) AP4, h) ENG, i) AP3, j) MS0.} 
\label{fig:l}
\end{figure*}

\begin{equation}
R_{m} = 2.9\times10^{8}\dot{M}_{17}^{-2/7}m_{NS}^{-1/7}\mu_{30}^{4/7} [cm]
\end{equation}

\noindent where $\dot{M}_{17}$ is the mass accretion rate at the truncation radius in units of 10$^{17}$ g/s, $m_{NS}$ is the neutron star mass in units of M$_{\odot}$ and $\mu_{30} = BR_{NS}^{3}/10^{30}$ G cm$^{3}$ (where $R_{NS}$ is the neutron star radius in units of cm). For magnetar field strengths (typically B $>$ 10$^{13}$ G) and a canonical neutron star radius of 10~km 
this might imply that the disc truncates before reaching the spherisation radius (see e.g. Mushtukov et al. 2015). However, in our Lense-Thirring model, we require that  $r_{m} < r_{sph}$ (where $r_m$ is $R_m$ in units of the gravitational radius) which would be broadly consistent with the sub-magnetar field strengths invoked by several authors (e.g. Kluzniak \& Lasota 2015; King \& Lasota 2016; Christodoulou et al. 2016) for ULPs observed to-date. In the specific case of P13, for the mass accretion rates we infer (based on a sensible mass range for the neutron star of $<$ 2.5 M$_{\odot}$) and based on the above formula for the magnetospheric radius, we restrict our analysis to field strengths $<$10$^{13}$~G to ensure $r_{m} < r_{sph}$ (although see also Israel et al. 2017b and the prospects of strong quadrupole field components but weaker dipole fields). 

In applying our technique to P13 - when the inner edge of the disc does not necessarily sit at the ISCO - we set $r_{in} = r_{m}$ in equation 5. We then determine $\dot{M}$ at $r_{m}$ by fixing the mass accretion rate at the ISCO to be Eddington limited (even though this radius is not actually reached by the inflow due to truncation). This approach makes the fundamental assumption that the position of $r_{m}$ (determined from equation 7) does not deviate substantially when the disc is thick (as it must be for $r < r_{sph}$) - this assumption is as yet untested but given the increasing interest in simulating this regime of accretion onto neutron stars, we can hope to gain a better understanding in the near future. In Table 2 we provide example parameter values from our model and in Figure 2 we show an example of the precession period as a function of mass and $a_{*}$ assuming a dipole field strength of 1$\times$10$^{12}$~G to be consistent with that inferred from spin-up (Fuerst et al. 2016) using the formulae of Ghosh \& Lamb (1979 - which we note assumes a thin disc geometry, unlikely to be present in these systems unless the field strength places $r_{sph} < r_m$). Although we can be fully certain that P13 contains a neutron star, in Figure 2 we also show the result for masses in the range of stellar mass black holes for illustrative purposes.


By comparing our predicted values for $P_{prec}$ to the {\it observed} period - and assuming this is due to Lense-Thirring precession - we then narrow down the possible mass-spin values for the neutron star in P13 which illustrates step iii) in \S2.
 
 \begin{figure*}
\begin{center}
\includegraphics[trim= 0 100 50 200, clip, width=16cm]{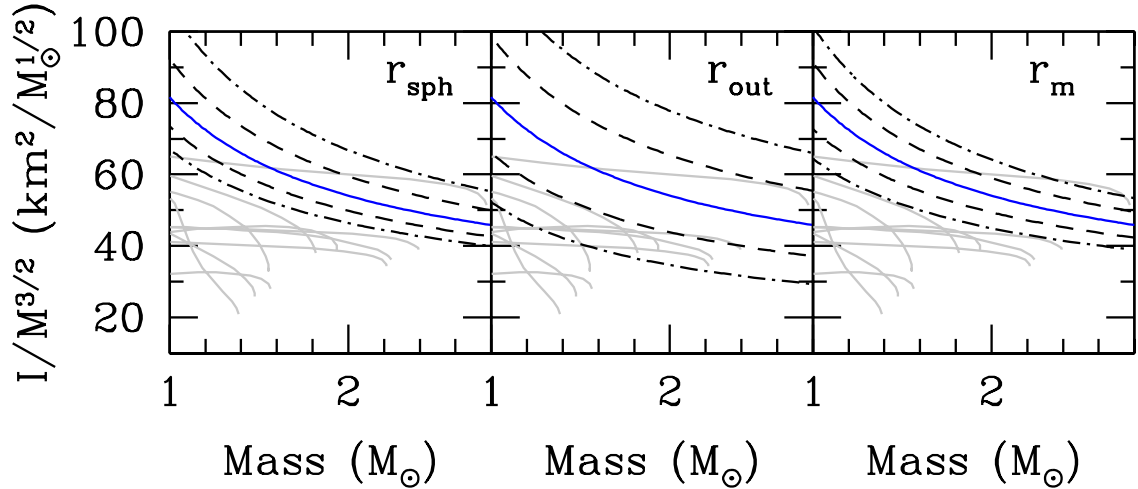}
\end{center}
\vspace{-0.2cm}
\caption{We plot the impact of 10\% (dashed) and 20\% (dot-dashed) systematic uncertainties on $r_{sph}$ (left-hand panel), $r_{out}$ (central panel) and $r_{m}$ (right-hand panel). In all three cases we overlay the resulting moment-of-inertia curves on the theoretical EoS (see Figure 3 for details) and assume B = 1$\times$10$^{12}$ G.}
\label{fig:l}
\end{figure*} 
 
\subsection{Moment-of-inertia/EoS constraints}

Following the steps outlined in \S2, we obtain values for the moment-of-inertia for a range of surface dipole field strengths (see arguments above) versus a range in neutron star mass; these are shown in Figure 3 overlaid onto a set of theoretical EoS (from Lattimer \& Schutz 2005). As might be expected from inspection of equations 5 \& 6, higher field strengths lead to stiffer EoS whilst lower field strengths lead to increasingly soft EoS. 

The 1 $\sigma$ statistical errors shown in Figure 3 result from the uncertainties on $T_{sph}$ propagated into our limit on $f_{col}$, $\dot{m}$, and subsequently $r_{sph}$ and $r_{out}$. Although we do not consider errors on the measured $P_{prec}$ (we use the best-fitting result obtained by Motch et al. 2014), from Figure 2 it is clear that for such long periods, the uncertainty in $a_*$ (and therefore $I$) is small even for large uncertainties in $P_{prec}$ of several days (see \S4.4 for a more detailed demonstration). 

It is important to note that there is no {\it a-priori} expectation for our technique to yield sensible moment-of-inertia values which could {\it potentially} be seen as an argument for Lense-Thirring precession driving the super-orbital period (although we consider alternative origins in \S5). Whilst these constraints on the EoS rely on a number of key assumptions, as we discuss in the following section, these can be addressed in future.

\section{Major sources of uncertainty}

Whilst we can estimate or directly measure the parameter values entering into the formulae for Lense-Thirring precession (\S2), we must also explore the effect of known statistical errors and potentially unknown systematics. In the case of the parameters entering into the super-critical disc model (\S 4.1-4.7), the uncertainty is {\it independent} of dipole field strength and we provide examples for B = 1$\times$10$^{12}$ G (once again noting that this estimate originates from spin-up assuming a thin disc geometry). Conversely, the impact of additional sources of torque from the secondary star and possible precession of the neutron star's magnetic dipole (\S 4.8) are both functions of the field strength and we take care to account for this in our discussion of the uncertainties.

\subsection{$r_{sph}$ and $r_{out}$}

Poutanen et al. (2007) determine that their formula for $r_{sph}$ (equation 1) is accurate to within 2\% of their numerical calculations. However, this formula does not account for unknown systematic errors and we explore the impact on the moment-of-inertia values of a 10\% and 20\% uncertainty on $r_{sph}$ and $r_{out}$. The result is shown in Figure 4; whilst the implied error is clearly much smaller in the case of $r_{sph}$ than for $r_{out}$, for lower dipole field strengths where the solution lies amongst many families of EoS, we may still struggle to differentiate between them for a 10\% uncertainty in either radius. However, in future we can hope to improve on the accuracy of these radial positions through the combination of broad-band (UV to X-ray) spectroscopy and utilisation of post-processed spectral models from 3D RMHD simulations (e.g. Narayan et al. 2017).

\subsection{$r_{m}$}

In the preceding sections we have used the formula for $r_{m}$ assuming a thin disc solution for the flow (Davidson \& Ostriker 1973); this is likely to be inaccurate given the accretion rates inferred for these objects and the abundance of radiation pressure leading to a large scale-height flow within $r_{sph}$. The exact solution will require a full prescription for the vertical and azimuthal stresses in the MHD flow and is beyond the scope of this paper and so we investigate the impact of a 10\% and 20\% error on $r_{m}$ in Figure 4. We find that the uncertainty in the moment-of-inertia solutions are similar in magnitude to those for $r_{sph}$ (see above).

\begin{figure*}
\begin{center}
\includegraphics[trim=90 110 20 150, clip, width=12cm]{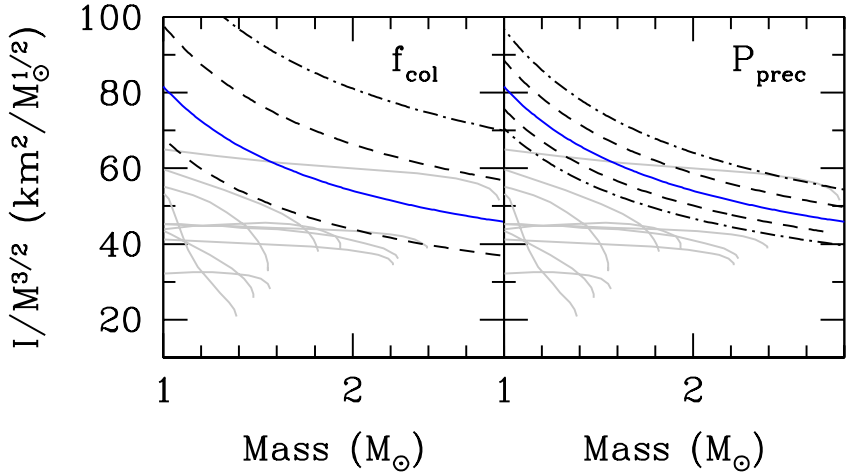}
\end{center}
\vspace{-0.2cm}
\caption{In the left-hand panel we show the impact of 5\% (dashed) and 10\% (dot-dashed) uncertainties on the limiting value of $f_{col}$, whilst in the right-hand panel we show the impact of uncertainties of 5 days (dashed) and 10 days (dot-dashed) in $P_{prec}$. In both cases we overlay the resulting moment-of-inertia curves on the theoretical EoS (see Figure 3 for details) and assume B = 1$\times$10$^{12}$ G.}
\label{fig:l}
\end{figure*}

\subsection{$f_{col}$}

As the temperature of the disc at $r_{sph}$ is above 10$^{5}$ K, we can be confident that the opacity of the inflow is dominated by electron scattering. However, the true value of $f_{col}$ is subject to substantial uncertainty; whilst the formula of Davis et al. (2006) reproduces consistent values even in the case of AGN (Ross et al. 1992; Done et al. 2012), any additional Comptonisation in the outflow (potentially due to turbulent motion, e.g. Kauffman, Blaes \& Hirose 2017) may lead to changes in the observed temperature and our inferred values for $\dot{m}$.

As increasing $f_{col}$ increases $\dot{m}$, the radii ($r_{sph}$ and $r_{out}$) also increase, resulting in a slower precession. As can be inferred from equation 5, to reach the {\it observed} precession period then requires a larger spin value and larger moment-of-inertia. As an example of the resulting uncertainty on the moment-of-inertia values, we show the impact of a 5\% (lower and upper) and 10\% (upper only, as we do not expect $f_{col} < 1.7$) error in our limiting value of $f_{col}$ in Figure 5. We note that at $f_{col} \gtrsim 2$ most EoS are excluded, however this may well be balanced by the impact of uncertainties in other parameters (e.g. $\epsilon_{wind}$ which we have also set at a limiting value - see \S 4.6).

Whilst we do not yet have a reliable estimate for $f_{col}$, through improvements in simulations of the accretion flow (e.g. Jiang et al. 2014) and by direct comparison to observations (e.g. Narayan et al. 2017) we can hope to obtain better constraints in the near-future. Finally, we note that any additional uncertainty on $T_{sph}$ (introduced by a different modelling of the continuum, e.g. Walton et al. 2017) is directly equivalent to an error in our limiting value of $f_{col}$.

\subsection{$P_{prec}$}

Although we have used a value for the precession period of P13 derived from observations in optical bands (Motch et al. 2014) and quoted without errors, we now investigate the change in the moment-of-inertia values for errors of 5 and 10 days on the observed super-orbital period, with the results plotted in Figure 5. Clearly, even for a large uncertainty in $P_{prec}$, the implied error on the moment-of-inertia is small as we would expect from inspection of Figure 2. We note that should we have instead used the recently reported X-ray and UV periods for P13 (both lying between 64-65 days with errors of $\sim$ 0.1 day: Hu et al. 2017) there would be little-to-no effect on the derived moment-of-inertia values.

\subsection{$v_{wind}$}

In determining the moment-of-inertia values we have assumed $v_{wind} \approx$ 0.1c to be broadly consistent with observations of canonical ULXs (see e.g. Pinto et al. 2017). However there is a range in the reported wind velocities (Pinto et al. 2016; 2017) with some closer to 0.2c; even in these cases, the observed velocity must be a {\it lower} limit on the true velocity unless we happen to observe the wind directly face-on. In Figure 6 we show the impact of increasing the outflow velocity to 0.15c and 0.2c; as expected, changing the velocity changes the value of $\psi$ which is  
equivalent to a fractional change in $r_{out}$ (see preceding sub-section). By inspection, a faster wind therefore results in a smaller $r_{out}$ and so a faster (smaller) precession period. To reach the larger {\it observed} period then requires a smaller spin value and lower moment-of-inertia whilst the converse is true for slower wind speeds. Whilst having a major impact on the EoS constraints, our estimate for $v_{wind}$ 
will no doubt be improved upon in follow-up work using higher energy-resolution spectroscopy (e.g. Pinto et al. 2016; 2017).

\begin{figure*}
\begin{center}
\includegraphics[trim=90 110 20 150, clip, width=12cm]{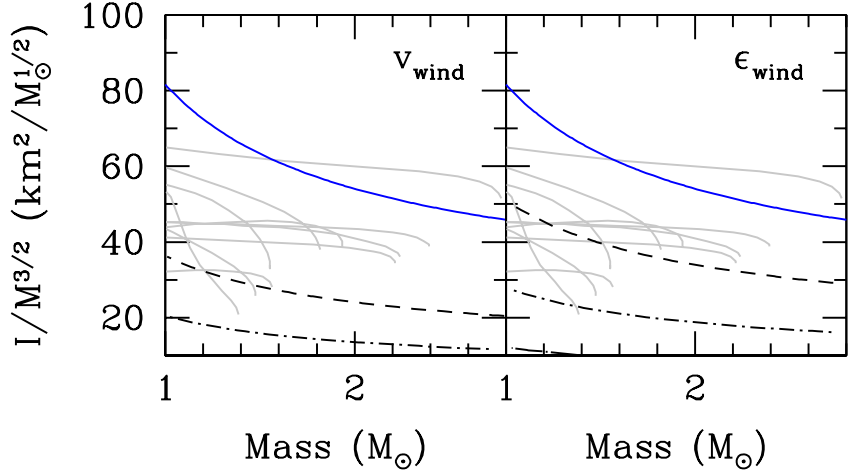}
\end{center}
\vspace{-0.2cm}
\caption{In the left-hand panel we show the impact of assuming wind velocities of 0.15c (dashed) and 0.2c (dot-dashed), whilst in the right-hand panel we show the impact of smaller values of $\epsilon_{wind}$ of 0.4 (dashed), 0.3 (dot-dashed) and 0.2 (dot-dot-dashed). In both cases we overlay the resulting moment-of-inertia curves on the theoretical EoS (see Figure 3 for details) and assume B = 1$\times$10$^{12}$ G.}
\label{fig:l}
\end{figure*}

\subsection{$\epsilon_{wind}$}

In the preceding sections we assumed a value for $\epsilon_{wind}$ at its approximate limiting value of 0.5 from simulations (S{\c a}dowski et al. 2014). At extremely high accretion rates ($\dot{m} \sim$ 1000), this value may increase substantially (Jiang et al. 2017), however, at rates consistent with our estimate for P13, $\epsilon_{wind}$ may well be $<$ 0.5 (Jiang et al. 2014) and in Figure 6 we show the impact of changing $\epsilon_{wind}$ to 0.2, 0.3 and 0.4 respectively. Clearly the resulting moment-of-inertia values are highly sensitive to this parameter. As with $v_{wind}$ we can hope to improve upon the estimate for $\epsilon_{wind}$ via high energy-resolution spectroscopy and detailed modelling.

\subsection{$\phi$}

We have assumed the opening angle of the wind-cone to be $\approx$ 55\degree~to be consistent with the results of simulations in this accretion regime (Jiang et al. 2014). We can determine the impact of uncertainties in this parameter from inspection; $r_{out} \propto 1/\phi$ where $\phi$ is the cotangent of the opening angle so a change of 5\degree~from our assumed value then corresponds to $\approx$ 20\% error in $r_{out}$ which we have explored above. The impact of this uncertainty may be mitigated via future constraints on the opening angle from studies of precession and direct modelling of the lightcurve (see Dauser et al. 2017).

\subsection{Additional torques}

Up until this point we have assumed that the Lense-Thirring torque is pre-dominant, however, there are undoubtedly competing stresses in the system and we now consider the impact on our EoS constraints by the most relevant of these. 
\bigskip

\subsubsection{Tidal torque from the secondary star}

Whilst the Lense-Thirring effect is an unavoidable consequence of general relativity for vertically misaligned orbiting particles, the size of the resultant torque for the small spin values we must have in slowly rotating neutron stars such as P13 is correspondingly small (Fragile et al. 2007) and, in principle, the torque from the tidal interaction of the secondary star could instead dominate. The ratio of the respective torques (Fragile et al. 2007) is given by:

\begin{equation}
\frac{\tau_{tidal}}{\tau_{LT}} = \left(\frac{m_{*}r_{sph}^{7/2}}{d^{2}}\right)\times\left(a_{*}m^{5/2}ln\left[\frac{r_{sph}}{r_{in}}\right]\right)^{-1}
\end{equation}

\noindent where $m_{*}$ is the mass of the companion star (in units of M$_{\odot}$) and $d$ is the binary separation (in units of gravitational radii). We note that, due to the dependence on $r_{in}$, this ratio is also a function of dipole field strength (i.e. when $r_{in} = r_{m}$). Whilst no binary period has yet been reported for P13 (we have good reason to suspect the 64-day period to be super-orbital - see \S3), supergiant HMXBs have periods of days (e.g. Corbet \& Krimm 2013 and as seen in M82 X-2: Bachetti et al. 2014), allowing the separation to be calculated according to (Frank, King \& Raine 2002):

\begin{equation}
d = 2.9\times10^{9}m^{1/3}(1+q)^{1/3}P_{d}^{2/3}\frac{c^{2}}{GM}  [R_{g}]
\end{equation}

\noindent where $q = m_{2}/m$ and $P_{d}$ is the binary period in units of days. For a sensible binary period range of 1-10 days and secondary mass estimated to be between 18-23 M$_{\odot}$ (Motch et al. 2014) we obtain the range in torque ratios shown in Figure 7 for B = 1$\times$10$^{12}$~G (once again assuming the field strength inferred from spin-up - Fuerst et al. 2016). Clearly the secondary star's influence is far less than that of Lense-Thirring for all masses and decreases with increasing orbital period (we also note that the impact decreases with decreasing field strength as $r_m$ decreases concordantly).
\bigskip 

\subsubsection{Magnetic torques}

It is important to consider the effect of the torque arising from the interaction between the neutron star's dipole field and the accretion disc when the neutron star's spin axis and magnetic axis are {\it both} misaligned with respect to the angular momentum vector of the binary orbit (see Lipunov \& Shakura 1980; Lai 1999). The former must occur for Lense-Thirring precession to take place whilst we know that the spin axis and magnetic axis are misaligned as we see pulsations; it is then not a huge leap for the magnetic axis to also be misaligned with that of the binary orbit. The discontinuity between magnetic field strengths above and below the disc plane leads to an induced radial surface current and the interaction of this with the external magnetic field from the neutron star then leads to a `magnetic torque' (Lai 1999; 2003). Where the magnetic torque dominates over the viscous torque, warping and precession is instigated (Lai 2003; Pfeiffer \& Lai 2004), the latter being retrograde with respect to the rotation of the neutron star (i.e. the magnetic torque is in opposition to the Lense-Thirring torque which leads to prograde precession with respect to the neutron star spin). It is crucial to note that simulations studying the role of magnetic torques have only been carried out in the thin-disc limit; appropriate simulations of a thick disc (as we must have for such high mass accretion rates) have not yet been performed. However, we can obtain the most conservative estimate of the impact from the ratio of magnetic torque to the Lense-Thirring torque (Lai 2003; Fragile et al. 2007) assuming the disc to be thin:

\begin{equation}
\frac{\tau_{B}}{\tau_{LT}} = \frac{B_{in}^2 \tan(\delta)r_{in}^6 (r_{out}^{-3} - r_{in}^{-3})} {48 \pi^2 a_* (GM)^{5/2} c^{-3} \Sigma_{in} \sin(\beta) r_{in}^{-1/2} \ln(r_{out}/r_{in})}
\end{equation}

\noindent where $B_{in}$ is the magnetic field strength near the inner edge of the precessing region (i.e. $r_{m}$), $\delta$ is the pitch angle of the magnetic field, $\Sigma_{in}$ is the surface density of the disc at the inner edge, and $\beta$ is the tilt of the disc with respect to the spin axis of the neutron star. The surface density at $r_{in}$ can be derived from considering the inflow to be Eddington limited at the ISCO and the subsequent scaling of $\dot{m} \propto r$ (Shakura \& Sunyaev 1973):

\begin{equation}
\Sigma_{in} = \frac{1}{\alpha} \frac{{\dot M}_{Edd}}{1000 r_{isco}} \sqrt{\frac{r_{in}}{GM}}
\end{equation}

According to standard theory, $B_{in} \approx (B_{NS}/2) \times (r_{NS}/r_{in})^{3}$ where $B_{NS}$ is the surface dipole field strength and the factor 1/2 arises from considering the equatorial rather than polar field. 
Assuming $ \tan(\delta)/\sin(\beta) \sim$ 1, $\eta$ = 1\% and $\alpha$ = 0.01 we then calculate the torque ratio for a range of dipole field strengths for canonical values of $M$ = 1.4 M$_{\odot}$, $R_{NS}$ = 10~km and $a_{*}$ = 0.001 (an appropriate value for this source - see Figure 2). The results are plotted in Figure 8 and indicate that, for field strengths above $\sim 10^{10}-10^{11}$~G, the flow is heavily influenced or dominated by magnetic torques which will act to slow or mitigate the Lense-Thirring precession. However, we stress that this is likely to be a {\it highly} conservative picture as the induced surface current is a function of the scale-height of the disc (for a fixed magnetic field change across the 
disc) and so we should expect a decrease in the effective magnetic torque by a factor $\sim (H_{thin}/H_{thick})|_{R}$ (i.e. the ratio of the heights of the thin and thick disc respectively at the same radius) which implies a potential decrease of more than two orders of magnitude (thereby allowing Lense-Thirring torques to dominate for higher surface dipole field strengths). We therefore expect that future 3D RMHD simulations exploring the induced current and torque will show that the magnetic torque is heavily diminished in the case of a super-critical disc.

\section{Alternative origins for precession}

Super-orbital periods on the timescales of tens to hundreds of days have been reported for a number of Galactic HMXBs (see e.g. Corbet \& Krim 2013) with a clear correlation between orbital and super-orbital period. Should ULPs be fed via Roche Lobe overflow (as expected in the case of the thermal expansion phase of a He core star), then the approximate super-orbital periods we see are a good match to those of other HMXBs. If on the other hand, ULPs are efficiently fed via a powerful wind, filling the companion's Roche lobe - which could be the case given the identification of supergiant companion stars with a number of ULXs (Heida et al. 2016) and the expected high mass loss rates from such stars ($>$10$^{-4}$~M$_{\odot}$/year: Matsuura et al. 2016) - then the observed super-orbital periods would be outliers when compared to wind-fed Galactic systems, potentially implying a different origin.

As discussed in Kotze \& Charles (2012), there are a number of alternative mechanisms to generate super-orbital periods in HMXBs, the most relevant of which we now consider as an alternative origin for the $\sim$64 day period in P13. 

\begin{figure}
\begin{center}
\includegraphics[trim=80 110 30 20, clip, width=8cm]{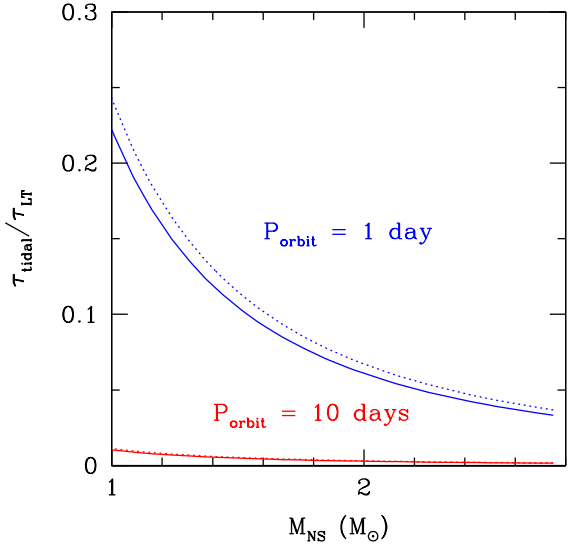}
\end{center}
\vspace{-0.2cm}
\caption{The ratio of the respective tidal to Lense-Thirring torques assuming a mass for the secondary star in the range of 18 (solid line) - 23 (dotted line) M$_{\odot}$ (Motch et al. 2014) and an orbital period consistent with observations of supergiant HMXBs (i.e. of order days, e.g. Corbett \& Krimm 2013). Whilst we have assumed B = 1$\times$10$^{12}$ G, we note that the torque ratio decreases with decreasing field strength. Clearly the Lense-Thirring torque is expected to dominate the flow.}
\label{fig:l}
\end{figure}

\subsection{Radiative warping}

A well-established theoretical means of warping the accretion disc is via the back pressure of radiation, emitted following illumination by a central source (Pringle 1996). Such non-linear, radiative warps can have a major impact on the structure of the disc at large radii which must then propagate inwards on a viscous timescale to misalign the entire flow. Ogilvie \& Dubus (2001) determine regions of stability and instability for such warps (based on the binary mass ratio and binary separation) which are well matched to observations of the super-orbital periods of a number of HMXB systems (e.g. Kotze \& Charles 2012), most notably Her X-1's 35 day period (Petterson 1977).

Based on the instability criterion presented in Pringle (1996) we can determine where a super-critical disc should be subject to radiative warps from the ratio of vertical to azimuthal viscosity (noting that this ignores the back pressure from any additional thermally-induced wind launching, which acts to reduce the warp radius: Pringle 1996). 
Figure 9 shows the ratio of Maxwell stresses in the polar and azimuthal directions from the simulations of Jiang et al. (2014). In this simulation, the authors set the primary to be a black hole with M $\approx$ 7 M$_{\odot}$ and $\dot{m}$ = 40; although the simulation does not account for the presence of a neutron star, we do not envisage a substantial change in the viscous stresses in the flow (assuming $r > r_{m}$). The simulation captures all of the expected physics of the flow including magnetic buoyancy (Socrates \& Davis 2006; Blaes et al. 2011; Jiang et al. 2013) which leads to vertical advection of radiation. Clearly the ratio of stresses shown in Figure 9 is far less than unity which could potentially place the warp within the spherisation radius. However, the 3D RMHD simulations never encounter a warp - an obvious reason for this is that incident flux will likely encounter the optically thick outflow rather than the underlying inflow; when this occurs, any reprocessed emission is advected away with the outflow and so there is no re-emission and back-pressure exerted on the inflow itself. Beyond $r_{sph}$, the disc is expected to be classically geometrically thin (Poutanen et al. 2007) and so we expect would be subject to radiative warps for viscosity ratios of order unity (Pringle 1996). However, the most luminous, inner regions are shielded from view by the large scale-height inflow/outflow. In addition, the radiative flux is advected with the outflow and simulations of the photosphere show that little of the radiation makes it to the outer disc - it would therefore seem that radiative warps are unlikely when the disc is classically super-critical (although we note that simulations exploring the stability of the outer disc in this regime of accretion are yet to be performed). 


\begin{figure}
\begin{center}
\includegraphics[trim=20 50 50 50, clip, width=8cm]{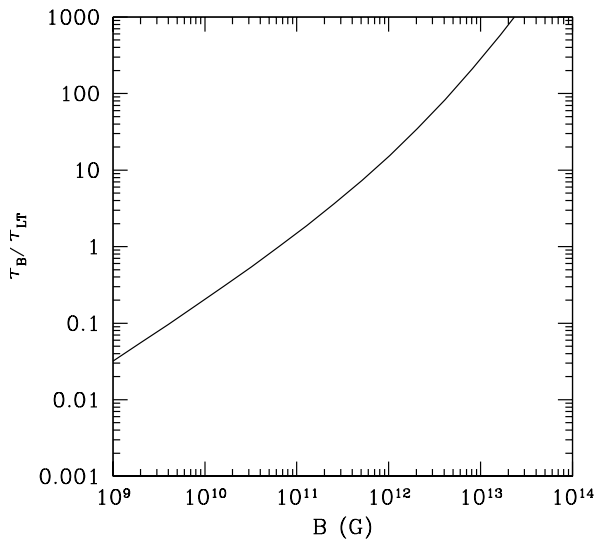}
\end{center}
\vspace{-0.2cm}
\caption{Ratio of magnetic torque (due to the interaction of the neutron star's magnetic field with the disc - Lipunov \& Shakura 1980) to the Lense-Thirring torque as a function of dipole field strength. Whilst it would appear that above a few $\times$10$^{10}$ G magnetic torques will dominate, we stress that this effect has only been studied in the thin disc regime which differs significantly from the super-critical regime and we expect the magnetic torque to be heavily diluted (see main text).}
\label{fig:l}
\end{figure}

Should the neutron star possess a strong enough dipole field such that $r_{m} > r_{sph}$ then radiative warps are entirely possible if not inevitable. In this situation, an accretion column is expected to have a fan-beam geometry or an accretion envelope is created within $r_{m}$ (which we note can also explain the smooth pulse profile seen in ULPs and may be able to recreate the thermal spectrum at high energies: Mushtukov et al. 2017); both of these scenarios lead to a luminous, vertically extended inner region which can then irradiate the outer regions of the disc leading to a warp. The precession timescales of this warp may then be similar to those seen in other Roche lobe overflowing HMXBs (although as we noted previously, this may not necessarily be the feeding mechanism).

\subsection{Magnetic precession}

As presented in Lipunov \& Shakura (1980), precession of the neutron star (and its dipole field) can lead to super-orbital variability that depends sensitively on the field strength (see also \S4.8.2). As discussed in Mushtukov et al. (2017), very high field strengths ($>$10$^{14}$ G) can lead to precession periods $\sim$months-year which has the attraction that it can self-consistently explain other features where very high field strengths are invoked - although see \S6 for a discussion on the uncertainty on the magnetic field strength in P13 (and in ULPs as a source class).

\subsection{Disc precession}

Resonance between particle orbits in the disc and the orbit of the secondary star can lead to precession of the disc and readily explain the presence of super-humps in the lightcurves of certain cataclysmic variables (Warner et al. 1995). However, this mechanism requires $q < 0.33$ (Whitehurst \& King 1991) which is highly unlikely to be the case in ULPs where the donor is expected to be high mass (e.g. Motch et al. 2014).

\begin{figure}
\begin{center}
\includegraphics[trim=120 50 120 50, clip, width=8cm]{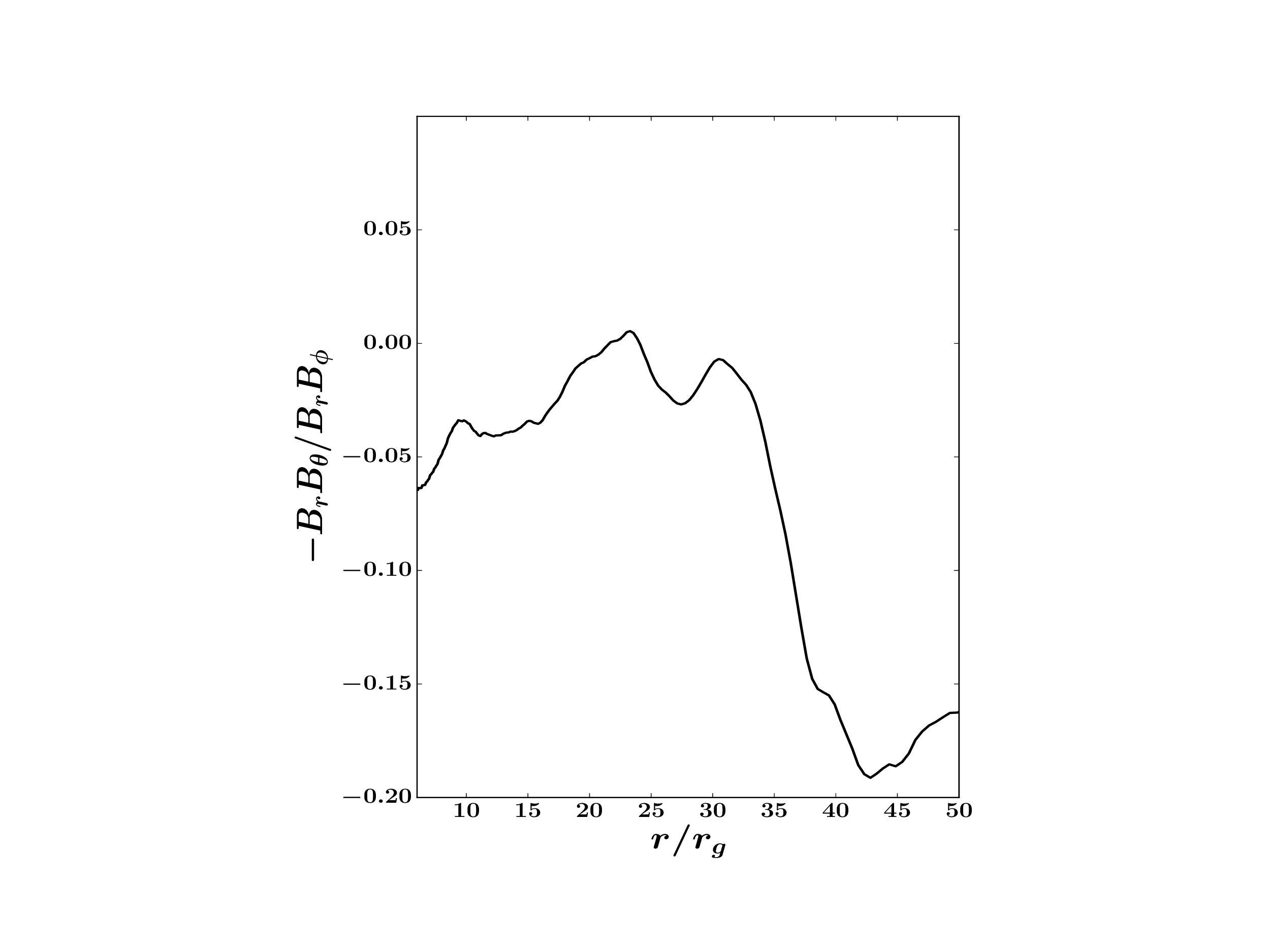}
\end{center}
\vspace{-0.2cm}
\caption{Ratio of the vertical (polar) and azimuthal components of Maxwell stress from the  3D RMHD simulation of Jiang et al. (2014). Clearly the ratio is far less than unity at all radii which could potentially place the warp within the spherisation radius, yet simulations have yet to encounter such a warp.}
\label{fig:l}
\end{figure}

\section{Discussion}

The prospect of determining the neutron star EoS from easy to isolate, long timescale trends in ULP lightcurves is naturally compelling, yet clearly there are several sources of uncertainty which may influence the outcome or mitigate the Lense-Thirring torque altogether. However, in future we can realistically hope to address some of the uncertainties; we may better localise the photospheric radius ($r_{out}$) from studies in the optical-UV band (where the peak in flux is at higher energies than the companion star), the spherisation radius ($r_{sph}$) and an estimate for $f_{col}$ from X-ray data. Vital to making the necessary progress is the application of physically motivated, post-processed spectral models from 3D RMHD simulations (e.g. Narayan et al. 2017) providing they subtend a sufficiently large parameter space. As is clear from \S4.5 and \S4.6, one of the major potential sources of error in our technique comes from our lack of tight constraints on $v_{wind}$ and $\epsilon_{wind}$; we can expect to improve on estimates for these parameters from high energy-resolution studies (see Pinto et al. 2016; 2017), especially with the advent of such facilities as {\it XARM} and {\it ATHENA}. It is interesting to note however that should our present estimates for these parameter values be close to correct, the current estimate of a moderate dipole field strength in NGC 7793 P13 ($\sim$ 1$\times$10$^{12}$ G: Fuerst et al. 2016) would yield stiffer EoS, consistent with the discovery of high mass neutron stars ($\approx$ 2 $M_{\odot}$) in milli-second pulsars (e.g. Demorest et al. 2010) which disfavours the softer EoS. 

Clearly one of the outstanding issues in interpreting the behaviour of ULPs (and the applicability of this technique) remains the as-yet unknown surface dipole field strength (see the discussions of King et al. 2017). In our analysis we have chosen to focus on sub-magnetar field strengths ($<$ 10$^{13}$ G), consistent with the findings of multiple authors across the three identified ULPs to-date (e.g. Kluzniak \& Lasota 2015; King \& Lasota 2016; Christodoulou et al. 2016; Fuerst et al. 2016). In addition, Walton et al. (2017) show explicitly the phase-averaged spectrum of P13 which appears consistent with that of other non-pulsed ULXs (see also Motch et al. 2014; Pintore et al. 2017) whilst we have tentative evidence for the presence of a radiatively driven outflow as seen in archetypal ULXs (Middleton et al. 2015b) - P13 would therefore appear to show all the hallmarks of a canonical ULX. Consistent with these observations, it has been suggested that a large component of the ULX population could be composed of super-critically accreting neutron stars (see King et al. 2001; 2016; Middleton \& King 2017) which then prompts us to ask whether the ULX class {\it as a whole} could be consistent with a picture where the super-Eddington luminosities arise from a pencil/fan beam/accretion curtain geometry and very high dipole field strength neutron stars. There are certainly compelling reasons to consider extremely high field strengths, notably the possibility of forming the emergent spectrum via an accretion curtain (see Mushtukov et al. 2017) and the possibility of precession via the motion of the neutron star's dipole field (e.g. Lipunov \& Shakura 1980) or a radiative warp (Pringle 1996).

Where accretion is taking place onto a high dipole field strength neutron star, the soft X-ray emission is associated with the accretion disc, peaking at $r_{m}$ and the hard X-rays within this radius (i.e. from the accretion column/curtain). The variability associated with the flow is presumed to be a consequence of local (MRI induced) turbulence and viscous propagation and is observed to emerge from the accretion cap (e.g. Uttley 2004). The limiting timescale is associated with the position of $r_{m}$ -- a larger truncation radius limits the available variability timescales to lower frequencies and reduces the sum total rms in the observable bandpass. It is well known that in ULXs the most variable sources are those in which the vast majority of the source flux emerges in the soft X-ray band (see Middleton et al. 2011, 2015a; Sutton et al. 2013) whist the variability is suppressed in the spectrally harder ULXs (see also Heil, Vaughan \& Roberts 2009). In order to satisfy this observational criterion, the accretion disc in a highly magnetised neutron star would need to move to smaller radii -- thereby allowing for more variability to enter our observable frequency range -- this would push the temperature of the soft X-ray component to higher temperatures. However, the most variable ULXs have the coldest soft components (see Middleton et al. 2015a) and so cannot be immediately reconciled within this picture. Conversely the coupled spectral-timing properties of classical super-critical flows can fully account for this behaviour (Middleton et al. 2015a). We therefore have additional reason to believe that $r_{m} < r_{sph}$ (which in turn would imply sub-magnetar dipole field strengths) although we stress that only an unambiguous measure of the field strength (via cyclotron lines - although higher multipole fields may influence such features) or an independent estimate of the truncation radius (e.g. via Fe K$_{\alpha}$ emission lines, although such features have yet to be detected in ULXs - see Walton et al. 2013) will solve this issue.

Irrespective of the uncertainty on the surface dipole field strength, it is clear that magnetic torques may result from the interaction of the neutron star's field with the disc which may in-turn lead to dilution of the Lense-Thirring torque. As we have noted, the impact of this effect - based on thin-disc formulae - is likely to be highly overestimated as the disc structure differs substantially when super-critical. Given the flurry of activity in simulating the accretion flow in this new object class, we can realistically hope to understand the role of torques in the disc more fully in the near future.

\section{Conclusions}

Whilst we cannot yet say with certainty what drives the super-orbital periodicity in ULPs, direct constraints on the surface dipole field strength will help resolve this issue, with weaker magnetic fields likely supporting Lense-Thirring precession. Although we cannot rule out a scenario where a radiative warp is generated in the outer disc, this would seem to demand that $r_{sph} <  r_{m}$ which would require extremely high (magnetar) surface dipole field strengths which can explain much of the details of ULPs including their spectra and smooth pulse profile (Dall'Osso 2016; Mushtukov et al. 2017) yet is disputed by a number of authors (e.g. Kluzniak \& Lasota 2015; King \& Lasota 2016; Christodoulou et al. 2016). Assuming that a large component of the ULX population contains neutron stars (although these may not be pulsing: King et al. 2017; Middleton \& King 2017), the coupled spectral-variability would also appear to argue for a more classical super-critical flow (Middleton et al. 2015a). 

If the long, super-orbital periods measured
in ULPs are the result of Lense-Thirring precession, we have demonstrated how these
periods can be used to constrain the moment-of-inertia of neutron stars and thereby provide
information on the EoS of nuclear matter. Whilst there are a number of potential sources of uncertainty inherent in our approach, many of these can be 
addressed via improved observations and the introduction of post-processed spectral models from 3D RMHD simulations (see e.g. Narayan et al. 2017) and their application to new and existing data. It is likely that more ULPs will
be discovered in the near future (e.g. when {\it eROSITA} launches in 2018). Our technique, applied to these new sources,
can therefore yield complementary and independent EoS constraints alongside those obtained from
other forthcoming measurements made by LIGO/Virgo, {\it NICER}, and the Square Kilometre
Array (SKA).


\section{Acknowledgements}

The authors thank the anonymous referee. MJM appreciates support from an Ernest Rutherford STFC fellowship. WCGH acknowledges support from STFC in the UK. TPR acknowledges funding from STFC as part of the consolidated grants ST/L00075X/1 and ST/P000541/1. AI acknowledges support from NWO Veni grant 639.041.437. CP and ACF acknowledge support from ERC Advanced Grant number 340442. This work is based on observations obtained
with {\it XMM-Newton}, an ESA science mission with instruments and
contributions directly funded by ESA Member States and NASA.

\label{lastpage}

\end{document}